\documentclass[%
 aip,
 pop,%
 amsmath,amssymb,
 reprint,%
]{revtex4-1}

\usepackage{graphicx}
\usepackage{dcolumn}
\usepackage{bm}
\usepackage{color,soul}
\usepackage{morefloats}

\usepackage{subfig}
\usepackage[newcommands]{ragged2e}

\begin{document}

\preprint{AIP/123-QED}

\title{Influence of electron-neutral elastic collisions on the instability of an ion-contaminated cylindrical electron cloud: 2D3V PIC-with-MCC simulations\\}

\author{M.Sengupta}
\affiliation {Institute for Plasma Research, HBNI, Bhat, Gandhinagar 382428, India}

\author{R.Ganesh}
\affiliation {Institute for Plasma Research, HBNI, Bhat, Gandhinagar 382428, India}

\date{\today}


\begin{abstract}
This paper is a simulation based investigation of the effect of elastic collisions and effectively elastic-like excitation collisions between electrons and background neutrals on the dynamics of a cylindrically trapped electron cloud that also has an ion contaminant mixed in it. A cross section of the trapped non neutral cloud composed of electrons mixed uniformly with a fractional population of ions is loaded on a 2D PIC grid with the plasma in a state of unstable equilibrium due to differential rotation between the electron and the ion component. The electrons are also loaded with an axial velocity component, $v_z$ that mimics their bouncing motion between the electrostatic end plugs of a Penning-Malmberg trap. This $v_z$ loading facilitates 3D elastic and excitation collisions of the electrons with background neutrals under a MCC scheme. In the present set of numerical experiments, the electrons do not ionize the neutrals. This helps in separating out only the effect of non-ionizing collisions of electrons on the dynamics of the cloud. Simulations reveal that these non-ionizing collisions indirectly influence the ensuing collisionless ion resonance instability of the contaminated electron cloud by a feedback process. The collisional relaxation reduces the average density of the electron cloud and thereby increases the fractional density of the ions mixed in it. The dynamically changing electron density and fractional density of ions feed back on the ongoing ion-resonance (two-stream) instability between the two components of the nonneutral cloud and produce deviations in the paths of progression of the instability that are uncorrelated at different background gas pressures.  Effects of the collisions on the instability are evident from alteration in the growth rate and energetics of the instability caused by the presence of background neutrals as compared to a vacuum background. Further in order to understand if the non-ionizing collisions can independently be a cause of destabilization of an electron cloud, a second set of numerical experiments were performed with pure electron plasmas making non-ionizing collisions with different densities of background neutrals. These experiments reveal that the nature of potential energy extraction from the electron cloud by the non-ionizing collisions is not similar to the potential energy extraction of other destabilizing processes \textit{e.g.} a resistive wall instability. This difference in the energy extraction process renders these non-ionizing collisions incapable of independently triggering an instability of the cloud.

\end{abstract}

\maketitle

\section{\label{sec:level1}Introduction\protect\\}

Contamination of a cylindrically or toroidally trapped electron plasma by a fractional population of ions can, under certain conditions, disrupt the equilibrium of the trapped electron cloud. The related instability, known as the ion resonance instability, manifests itself in the form of growing poloidal modes (diocotron modes\cite{rdav1,dub}) of the plasma\cite{megh3}. The primary cause for such contamination (either planned or inadvertent) in most electron plasma trap experiments is the process of ionization of the neutral background gas in the trap by the electrons. A brief history of analytical modelling\cite{levy,rdav1, rdavh1,rdavh2,faj,rdavh3}, numerical analysis\cite{chen}, and experiments\cite{eck,per,bet,kab,bet2,mark,stone,lach} performed to understand this destabilization process can be found in Sec I of Reference 3\cite{megh3}. Of the theoretical models for the ion resonance instability, the one developed by Davidson and Uhm\cite{rdavh1} aptly describes the collisionless dynamics of the instability in the present set of numerical experiments performed with an initially cold, partially neutralized electron cloud in cylindrical confinement. The model describes the ion resonance instability as a rotating two-stream instability between the electron component and the partially neutralizing ion component of the plasma. The differential rotation between the components in mutual equilibrium drives the instability; the rotation itself being a result of the radial self-electric field of the nonneutral plasma. Davidson and Uhm's linearised model\cite{rdavh1} works for any cold equilibrium of the two component plasma in cylindrical confinement. It shows that depending on the equilibrium parameters, one or more exponentially growing diocotron modes may be excited on the plasma profile with the equilibrium condition determining the fastest growing diocotron mode in the system. An abridged version of the analytical procedures for this model can be found in Sec III of Reference 3\cite{megh3}. 

In an earlier work\cite{megh3} we performed numerical experiments to investigate the full nonlinear evolution of the ion resonance instability in partially neutralized electron clouds in cylindrical traps. The numerical experiments were performed for a set of different initial unstable equilibria of the two component plasma (defined by the equilibrium parameters), and a well benchmarked\cite{megh, megh2} 2D Electrostatic Particle-in-cell (PIC) code was utilized for these simulations. Results\cite{megh3} from the initial linear growth phase of the instability were in excellent agreement with the linearised analytical model of Davidson and Uhm\cite{rdav1} which served as an added benchmark for the code. The simulations delved further in time beyond the linear growth phase and revealed several new and interesting nonlinear phenomena associated with the instability. For example, a process of simultaneous wave breaking of the excited poloidal mode on the ion cloud and pinching of the poloidal perturbations on the electron cloud occurring in the nonlinear phase of the instability was reported\cite{megh3}. This simultaneous nonlinear dynamics of the two components was found to be associated with an energy transfer process from the electrons to the ions. At later stages of the nonlinear phase, a heating induced cross-field transport of the heavier ions and tearing across the pinches on the electron cloud followed by an inverse cascade of the torn sections was also demonstrated\cite{megh3}. 

These ion resonance instability simulations of the previous work\cite{megh3} were performed in an ideal scenario with a perfect vacuum as a background. For the present work we have taken these simulations a notch closer to a usual experimental scenario by including the effect of elastic and excitation collisions of the electrons with background neutrals that are invariably present at experimental low pressures inside the trap. The first set of numerical experiments of this paper (Sec III) are PIC-with-MCC simulations that investigate how collisional relaxation of the electron cloud's profile can influence dynamics of the ion resonance instability happening due to the ion contamination of the cloud.

In the second set of numerical experiments of this paper (Sec IV) we initiate electron clouds in stable $l=1$ ($l$ being the diocotron mode number) mode orbits of very small radius, and utilize the MCC to simulate non-ionizing elastic and excitation collisions of these electrons with a neutral background in course of the simulations. The objective of this set of numerical experiments is to test if such non-ionizing collisions can destabilize the stable motion of the pure electron clouds. It had been theorized by Davidson and Chao\cite{chao} that elastic collisions between electrons and background neutrals will not only cause relaxation of the electron cloud's profile but can also destabilize any small-amplitude, stable $l=1$ mode  azimuthal asymmetry present in the cloud into a growing $l=1$ mode by virtue of their capability of extracting potential energy from the cloud. Contrary to these theoretical predictions, our simulations show that such non-ionizing collisions between electrons and background neutrals can not by themselves destabilize a stable configuration of the cloud. We believe that this has got to do with the manner in which potentially energy of the cloud gets reduced by such collisions. A descriptive comparison of the potential energy extraction process between non-ionizing collisions and a resistive wall instability\cite{white,bet3} that is known to be capable of destabilizing the cloud, will make this point clear. This is presented in Sec IV of this paper.

A brief outline of the following Sections of this manuscript is as follows. Sec. II is a description of the 2D3v PIC code with newly incorporated MCC facility. The PIC-with-MCC simulations of partially neutralized electron clouds and initially axially asymmetric pure electron clouds will be presented in Sec. III and Sec. IV respectively. The numerical MCC scheme used to produce the results of Sec. III and IV is validated statistically and numerically in Sec. V. The MCC verification procedure is carried out for one of the experiments of Sec. III.  A summary of the simulation results and related discussions make up the last section of the paper, Sec. VI.

\section{The 2D3v PIC code with MCC facility}

Different technical aspects of our PIC-with-MCC code, and their signifance are described in the following sub-sections.

\subsection{From the 2D PIC code to the 2D3v version}

A previously developed 2D Electrostatic PIC has been upgraded to a 2D3v PIC code with facility for Monte-Carlo-Collisions for the purpose of simulating the collisional interactions of nonneutral plasmas with background neutrals along with the collisionless dynamics of the plasma in cylindrical trap cross-sections. The 2D PIC code\cite{megh3, megh, megh2} (earlier version) was developed in FORTRAN-90 using Cartesian co-ordinates, and parallelized with OPEN-MP. It can simulate cross sections of multi-component plasmas of varying neutrality, confined within any perfectly conducting closed boundary curve. The size, shape, and toroidal aspect ratio of the boundary can be manoeuvred as per requirements. The numerical schemes employed in the development of different parts of the code along with references to the sources\cite{bird,hal,ket,cz,lal,chin} of these numerical procedures is described Sec. II of Reference 3\cite{megh3}. The 2D code has passed rigorous benchmarking with analytical results for cylindrically trapped pure electron plasmas\cite{megh,megh2} as well as partially neutralized electron plasmas\cite{megh3}. The 2D3v PIC-with-MCC version of the code retains all the features of the original 2D code. In subsections B to D we will discuss only those new features of the 2D3v PIC-with-MCC version that have been applied for the numerical experiments of this paper. For convenience of describing these numerical features of the code we will be explicitly referring to the computational particles of simulation as pseudo-electrons/ions throughout Sec. II instead of loosely calling them electrons/ions as in the rest of the paper. The last two subsections, E and F, describe the setup of the numerical experiments of Sec. III-IV and the numerical diagnostics used in the simulations respectively.   

\subsection{Giving the pseudo-electrons an axial velocity component perpendicular to the 2D PIC plane}

In addition to the 2D Cartesian components of velocity the pseudo-electrons now have a $3^{rd}$ velocity component representing their axial velocities. A common fixed magnitude of this component is initially given to all pseudo-electrons. The axially up or down direction for this component is selected randomly using a pseudo random number generator. The axial velocity does not get modified by the 2D PIC dynamics of the pseudo-electron. It can get modified only when the pseudo-electron is scattered by a collisional event. Thus the extra axial component, with a suitable choice of magnitude, will have a very similar effect on the collisional dynamics of the pseudo-electrons in simulation, as the axial bouncing of electrons between electrostatic end plugs\cite{malm} have on the collisional dynamics of electrons in cylindrical trap experiments. 
\subsection {The MCC procedure}

The MCC is a technique of making computational particles (here pseudo-electrons) undergoing guided motion in a simulation collide at suitable intervals (collision time step) with background matter that is not represented in any form in the simulation (here neutral Argon atoms)\cite{vah}. The method is useful when we are interested only in the effect (through collisions) of the latter on the former and not vice versa. To set up the MCC within the framework of a PIC code, the computational particles (here pseudo-electrons) of the simulation have to be treated as real particles (here electrons) inside the MCC algorithm\cite{cbird}. This implies that mass of a pseudo particle entering the MCC routine gets divided by its representation number. MCC can be implemented using a pseudo random number generator if the collision cross-section of a particle (here electrons) is available in analytical or numerical form as a function of its kinetic energy\cite{frig}. Taking the average kinetic energy of a particle (particle here being a pseudo-electrons represented as  an electron by scaling down its mass) in the collision time step its collision probability in that interval is determined from the available analytical/numerical collision cross-section. Then the pseudo random number generator churns out a pseudo random number that decides whether the particle undergoes a collision or not based on its collision probability (the Monte-Carlo method). If there are more than one kind of collision possible between the particle and the background then the generated random number can also decide between different possible types of collision and the possibility of no collision for the particle\cite{nan2}. Of course the probability of each type of collision in the collision time step should be available\cite{frig}. For the present set of experiments only elastic collisions and the most abundant first level excitation collisions of background atom has been considered. 

If the fate of the particle is found to be no collision then its velocity components are left untouched else the code proceeds to the next step of the MCC which decides how the particle gets scattered by the collision. If the collision happens to be of the elastic type then using a set of numerical formulas relying on pseudo random numbers the scattered speed and direction of the particle from the collision is determined\cite{nan1,okh}. These formulas are fed with the incident kinetic energy of the colliding particle and also the fixed temperature and pressure of the background gas to produce the required outputs. In our MCC code the set of formulas for elastic collision execution are very realistic. They implement the change in kinetic energy of the particle as a result of the elastic collision as well as the natural anisotropy of the scattering with respect to the incident direction of the particle. The scattering angle is determined from a very accurate formula for differential scattering cross section introduced in the year 2002\cite{okh}. This formula is actually a correction on the differential scattering cross section formula that had been in use earlier\cite{vah, nan1}. The colliding neutral though massive (in comparison with electron) is not approximated as a stationary body in the collision but its inherent thermal velocity is also taken into account in the scattering procedure through pseudo random numbers\cite{nan1}. 

If on the other hand, the collision happens to be a first excitation of the background (Ar atom), then the MCC follows a two step algorithm to simulate the collision. First the kinetic energy of the particle (pseudo electron with mass scaled down to electron mass) is reduced by an amount equal to the first excitation energy (= 11.55 eV for Ar atom)\cite{vah}. Next the particle with reduced velocity goes through an elastic scattering procedure as explained above. The first-level excitation collisions have been grouped together with, and treated as elastic collisions in the numerical experiments of this paper. This is because there is an elastic collision happening in every exciting collision adding to the relaxation of the electron component profile which is the subject of interest here.
 
After the MCC procedure is complete the masses of the computational particles are again scaled up to their original values for the regular PIC part of the code.

The procedure adopted for the MCC in the PIC-with-MCC code can be summarized as follows:
\begin{itemize}
\item{Scale down the mass of the pseudo electron to that of an electron}
\item{Now calculate the electron's kinetic energy}
\item{Calculate the probability of elastic collision, probability first level excitation collision, and probability of no collision for that kinetic energy of electron}
\item{Generate a pseudo random number to decide the fate of the particle}
\item{a) For elastic collision use pseudo random number based formulas to decide kinetic energy after scattering and scattered direction of the electron b) For first level excitation reduce energy of the electron by a magnitude equal to first level excitation energy of the neutral and then use the elastic collision formulas to determine final scattered speed and direction c) For no collision leave the velocity components of the particle untouched}
\item{Scale up the mass of the particle to the representative mass of pseudo electron for the regular PIC part of the code}
\end{itemize}

\subsection {The collision time interval}
The interval of time representing the collision step is a critical parameter which determines how closely and how efficiently the MCC simulates the collisional interaction of particles with the background. The first criterion the collision time step should satisfy is that its value should be so small that possibility of simulation particles colliding more than once with the background matter in this interval is practically removed all through the simulation\cite{vah}. Otherwise the MCC routine will overlook possible collisions leading to divergence of the simulation from a real experiment. It can be calculated that a collision time step in which the highest (among particles) probability of a collision remains below the value $0.095$ will satisfy this first criterion\cite{vah}. 

At the same time one should avoid diminishing the size of the collision time step beyond the point at which average (over particles) probability of collision in the interval becomes less than $1/N$, where $N$ is the size of the colliding pseudo-particle population. This is because such a choice will result in too many executions of the MCC routine in the simulation period with zero or very few pseudo particles actually colliding in each execution. Hence it will bring down the speed of execution of the PIC-with-MCC code. However improving efficiency of the MCC by increasing size of the collision time step should only be attempted as long as the primary criterion (related to accuracy of simulation) is satisfied for all pseudo particles. 

In our simulations the pseudo electrons are loaded in rigid rotor equilibrium with a fixed angular velocity and a common fixed magnitude of the axial velocity component. Hence a pseudo electron lying at the half-radius of the cylindrical, uniform density electron cloud will have the mean speed and the average collision cross section for electrons in the cloud. Hence in order to ensure that the first criterion is satisfied for our chosen collision time step we checked that the collision probability of the particles at half-radius of the cloud is well below the $0.095$ limit\cite{vah}. Also the size of the collision time step was adjusted so that on a average a fraction of $1.197\times10^{-3}$ from the total pseudo electron population of $87834$ collides in every collision step, thus satisfying the second criterion as well.

\subsection{The numerical experimental setup and its PIC parameters}
All simulation experiments described in sections III and IV of this paper were carried out for typical parameters found in experiments of cylindrical traps\cite{malm,dub}, in particular, a wall radius, $R_w=0.125m$, and axial magnetic field, $B_z=0.015T$, and radial extent of the uniform density plasma, $r_p=0.5\times{R_w}$ . The simulation time step, ${\delta}t=10^{-11}sec$ is chosen much smaller than the cyclotron time period, $T_{ce}=2.38\times10^{-9}sec$ of electrons, such that the code can well resolve the cyclotron motion of both ions and electrons. The collision time step ${\Delta}t_c$  is adjusted according to the chosen pressure and temperature of the background gas. The magnitude of axial velocity with which the electrons are loaded is $1.02727\times10^7 m s^{-1}$ which implies that for a typical cylindrical plasma column of length of $60 cm$\cite{dris2} the bounce frequency of electrons is roughly $8.5 $ MHz. This value of the axial bounce frequency is higher than the typical experimental bounce frequency ($<0.2$ MHz\cite{dris2}) of a $60 cm$ long cylindrical plasma column. By loading the electrons with high axial speeds we have made the electrons more energetic and thereby maintain the average total non-ionizing (elastic + excitation) collision cross section between the moderate orders of $10^{-21}\,m^2$ and $10^{-20}\,m^2$ (see Fig. 3 of Reference 30\cite{vah}) in the numerical experiments of Sec III and IV. It must be re-emphasized here that the loaded value of the axial velocity of electrons in the simulations serves only to adjust the collisional probabilities and does not partake in any of the regular PIC dynamics of the plasma.   

PIC-parameters used in the simulations of Sec. III are $87834$ pseudo particles for each plasma component, on a $70\times70$ grid. As the actual population of the ions is a fraction, $f$ of the electron population, the number of real ions represented by each ion pseudo-particle has also been scaled down to a fraction, $f$ of the representation value of electron pseudo-particles. The ion species (H+ ion) is $1836$ times heavier than an electron  and the background atom (Ar atom) is $72820.77$ times heavier than an electron. 

In Sec. IV, the simulations are for pure electron plasmas represented by $87834$ pseudo electrons on a $70\times70$ grid. The background species is again Argon. 

\subsection{Diagnostics for the simulations}

The diagnostics that will be used in the experiments of Sec. III and IV are 4 potential probes which are basically 4 azimuthally equispaced Cartesian cell nodes inside the cylindrical trap that are all located at a common radius that is very close to the wall. As per their location on the trap cross-section the 4 potential probes have been named as left probe, right probe, top probe, and bottom probe.  The potential probes signal are used to study the dynamics of the plasma, and find the frequencies and growth rates of diocotron modes excited on the plasma\cite{megh3,megh}. Besides the potential probes the energetics (kinetic and potential energies versus time) of both components of the plasma, and the radius of the centre-of-mass of the electron component as a function of time have also been used as diagnostics in Sec. III and IV.

\section{Ion resonance instability in presence of electron-neutral elastic collisions}

For the numerical experiments of this Section we loaded two uniform density 2-component (e- and H+) nonneutral plasma profiles in their respective rigid rotor equilibrium at different background gas pressures. The first profile was loaded with a value of electron component Brillouin ratio, $f_b=0.02$ and a fractional neutralization, $f=0.15$ provided by the uniformly mixed H+ ions. The second profile had values $f_b=1.0$ and $f=0.1$. In both cases the electron and the ion components of the cylindrical, uniform density nonneutral plasma are rotating in the slow mode of the mutual rigid rotor equilibrium\cite{megh3}. Hence the electron component is rotating with angular velocity $\omega_{re}^{-}$ while the ion component is rotating with angular velocity $\omega_{ri}^{-}$ in mutual equilibrium\cite{megh3}.

\begin{figure}
\centering
\includegraphics[scale=0.7]{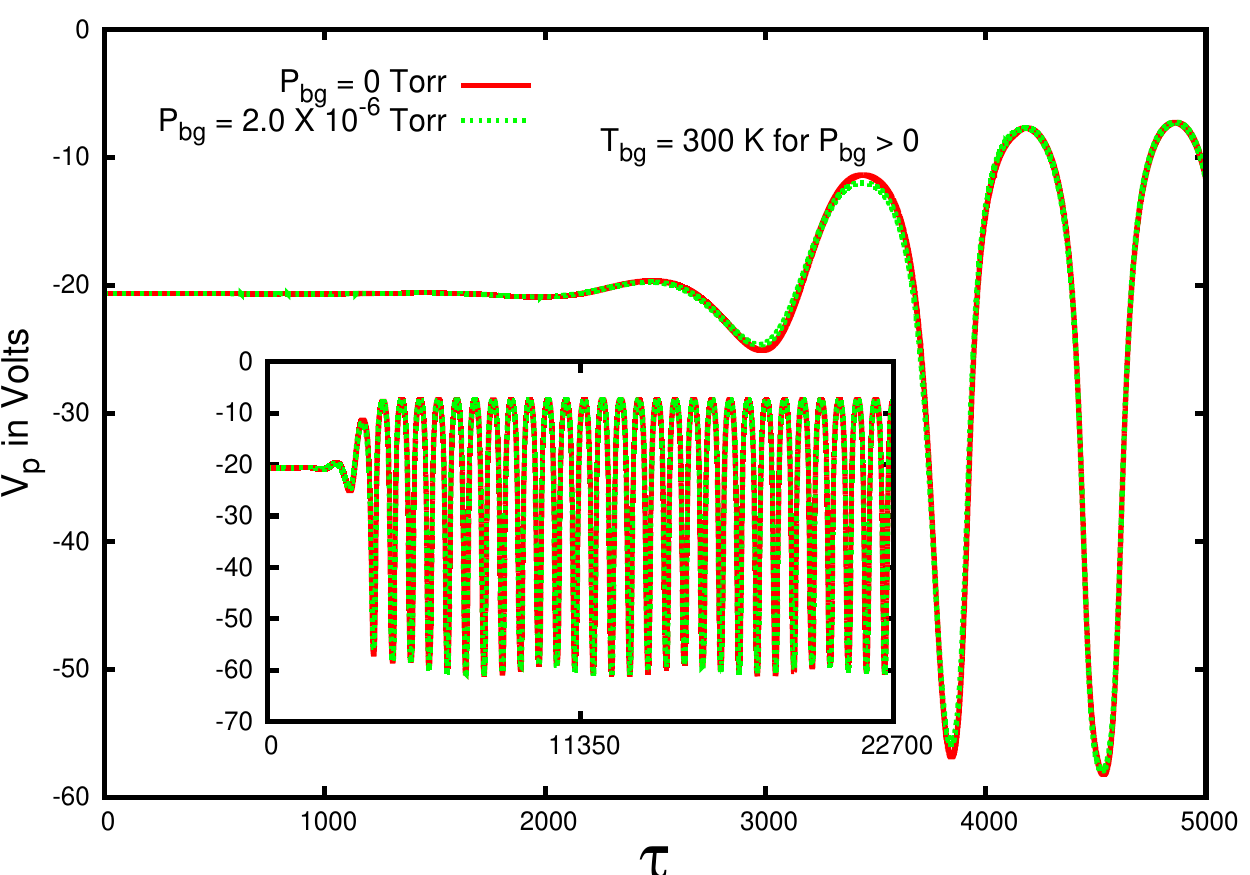}
\captionsetup{justification=raggedright,
singlelinecheck=false
}
  \caption{Readings of the left potential probe, $V_p$ for the $f_b=0.02$, $f=0.15$ equilibrium at background pressure, $P_{bg}$ values of $0$ and $2\times10^{-6} Torr$. $V_p$ readings of the equilibrium at the other two simulated background pressures of $2\times10^{-7} Torr$ and $2\times10^{-8} Torr$  (not shown here) also overlap nearly perfectly with the plotted readings in this figure. $T_{bg}=300 K$ is the chosen temperature of the background gas in this set of runs for all $P_{bg}>0$. Normalized time, $\tau$ is in units of electron cyclotron time, \textit{i.e.} $\tau=t/T_{ce}$. Readings upto the growth phase (till $\tau=5000$) have been zoomed in here for clarity while the inset has the complete readings upto the end of the simulations. The other three potential probes also recorded almost perfectly overlapping readings for all the four values of $P_{bg}$.} 
\end{figure} 

\begin{figure}
\centering
  
\includegraphics[scale=1.0]{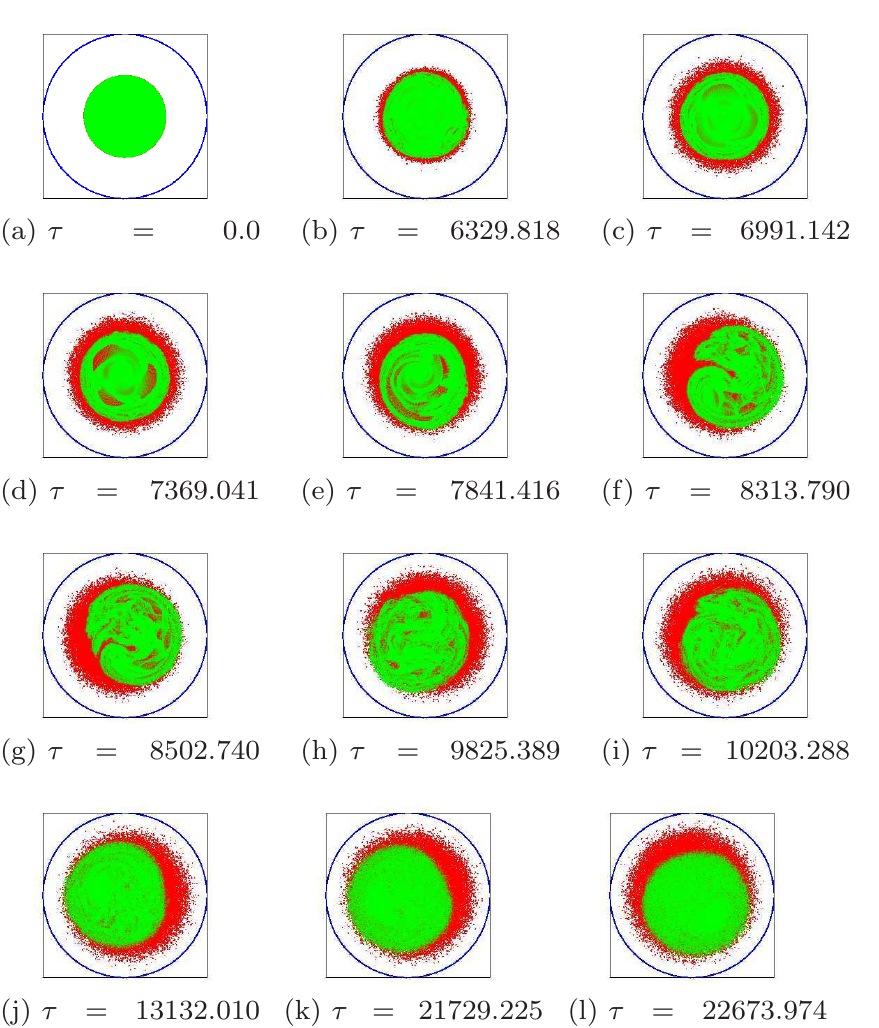}

\captionsetup{justification=raggedright,
singlelinecheck=false
}
  \caption{Snapshots of pseudo particles (electrons in red and ions in green) for the $f_b=1.0$, $f=0.1$ equilibrium load at zero pressure ($P_{bg}=0$) that excites a nonlinear $l=1$ mode . Below each snap, the time elapsed is mentioned in normalised units of electron cyclotron time, \textit{i.e.} $\tau=t/T_{ce}$.}
\end{figure}

\begin{figure*}
\includegraphics[scale=1.0]{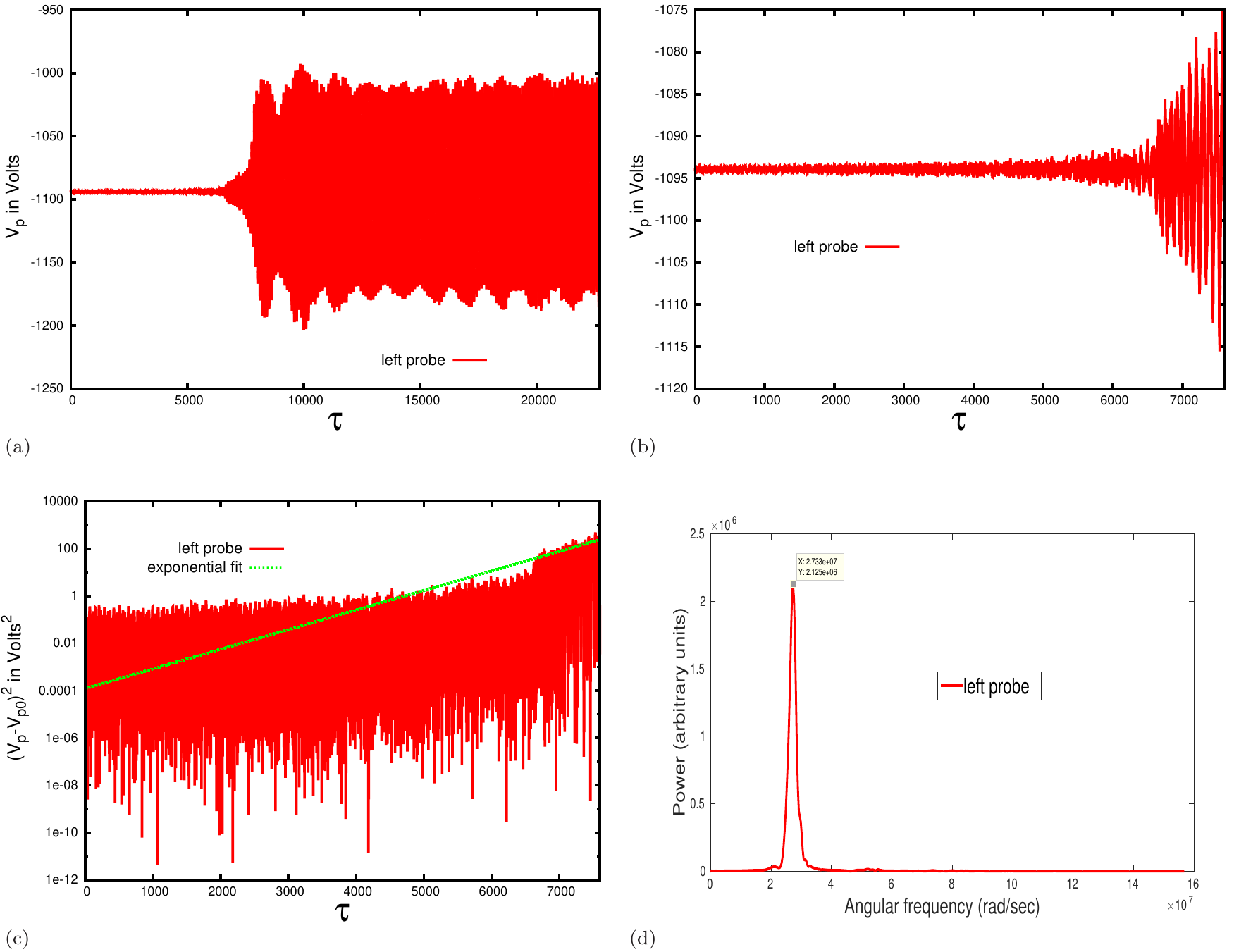}   
\captionsetup{justification=raggedright,
singlelinecheck=false
}
\caption{Potential-probe signal analysis of the $f_b=1.0$, $f=0.1$ equilibrium load at zero pressure ($P_{bg}=0$). The analysis of the left probe's signal is shown here. The other three potential probes also give the same results from their data analysis: (a) $V_p$ is the electrostatic potential recorded by the left probe. Normalized time, $\tau$ is in units of electron cyclotron time, \textit{i.e.} $\tau=t/T_{ce}$. (b) is a zoomed in plot of the same signal upto $\tau=7600$. This part of the signal has been exponentially fitted for a linear approximation of the growth of the $l=1$ mode, and has been Fourier analysed for an estimate of the mode's frequency. (c) is an exponential fit on the readings of (b). $V_{p0}$ is the left probe's reading at $\tau=0$ . The y-axis is in log scale while time axis is in linear scale. (d) A FFT is performed on the readings of (b). Here the x-axis is in rad/sec unit and the y-axis is the power factor in arbitrary unit.}
\end{figure*}

\begin{figure}
\centering
\includegraphics[scale=0.7]{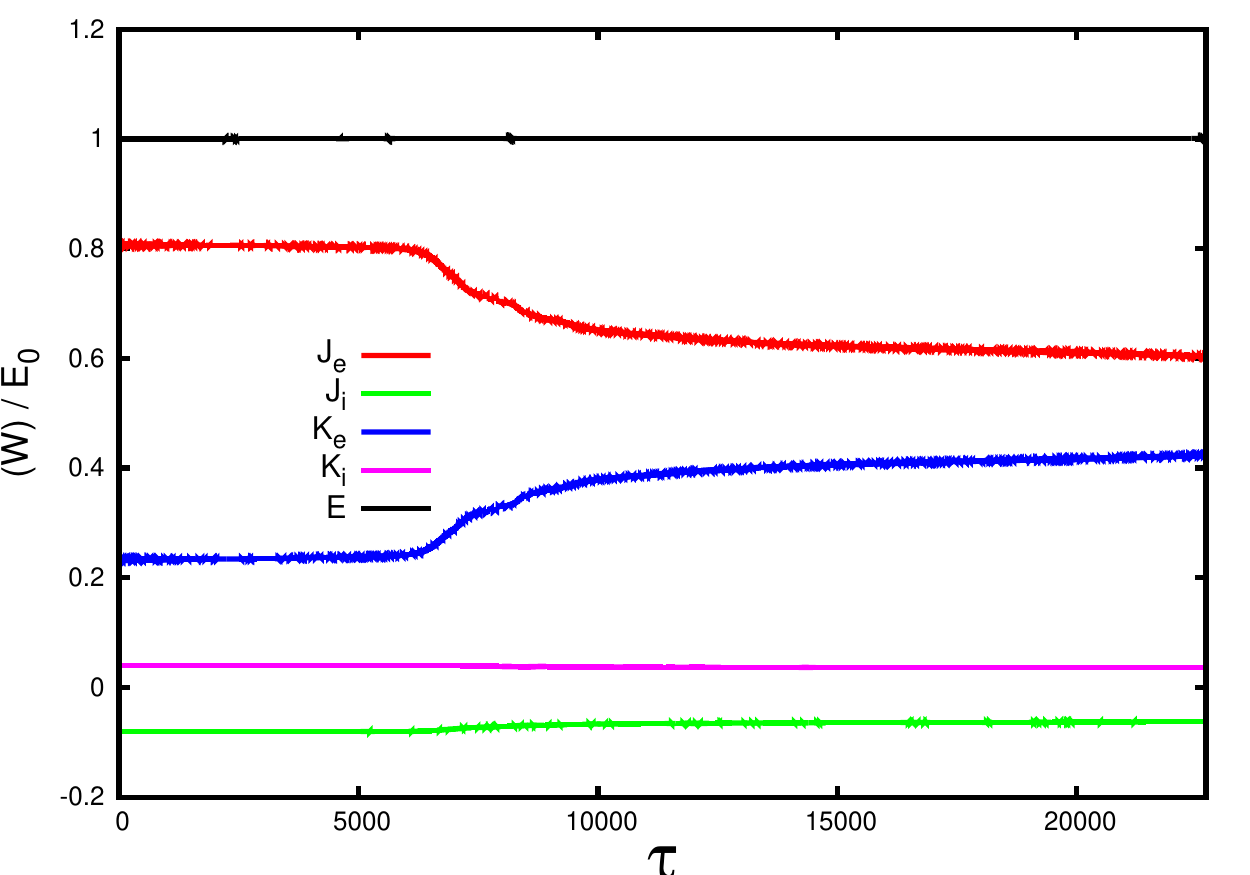}
\captionsetup{justification=raggedright,
singlelinecheck=false
}
  \caption{Energetics of the $f_b=1.0$, $f=0.1$ equilibrium load at zero pressure ($P_{bg}=0$): $W$ represents the energy components in the legend, $E_0$ is the initial total energy, and $\tau=t/T_{ce}$. $J_e$ and $J_i$ are the potential energies of electrons and ions respectively. $K_e$ and $K_i$ are  the kinetic energies of electrons and ions respectively. $E$ is the total energy of the system. The energy components are normalised by $E_0$ and plotted as a function of time. The time axis is normalised by the cyclotron time period of electrons, $T_{ce}$.}
\end{figure} 

\begin{figure*}
\includegraphics[scale=1.0]{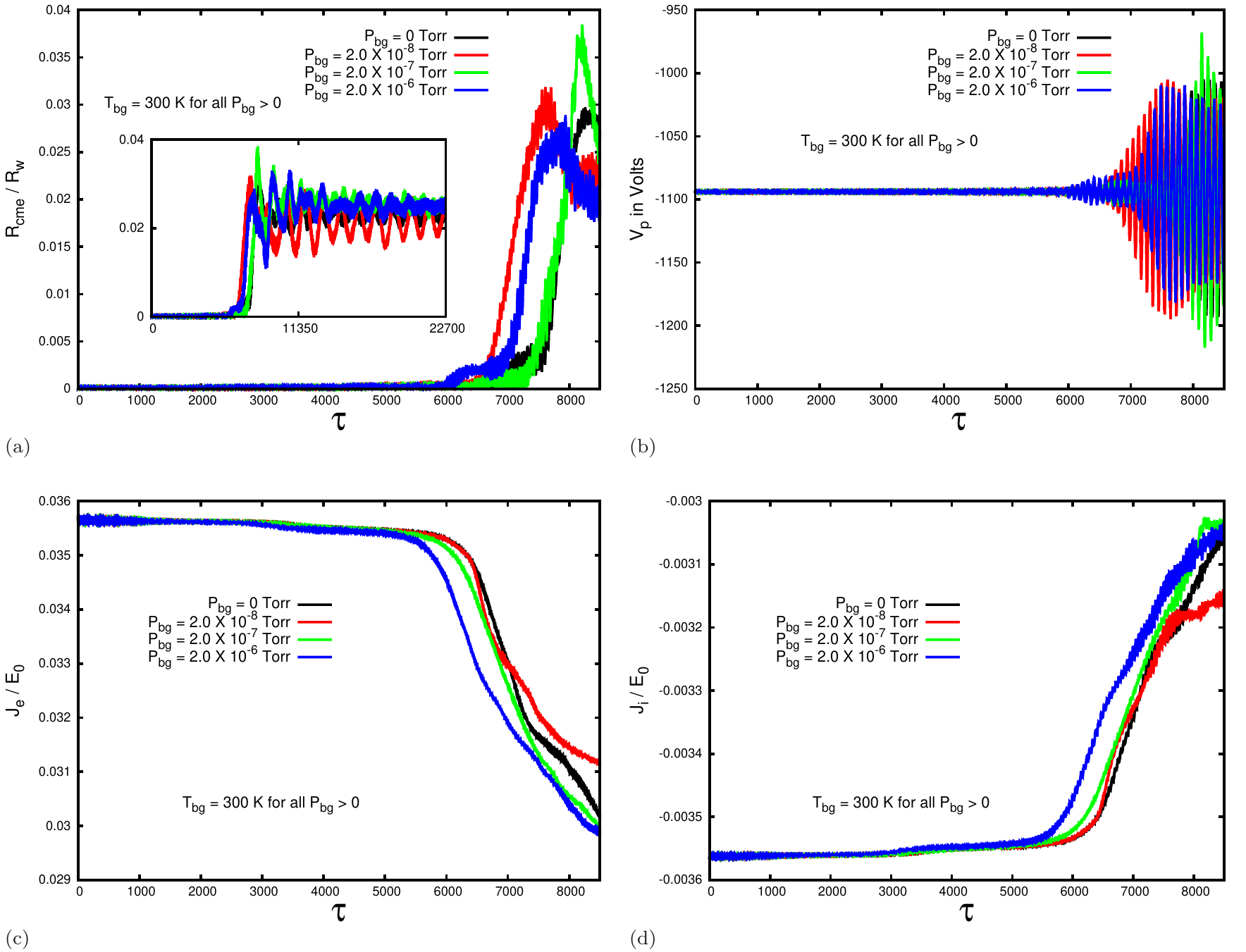}   
\captionsetup{justification=raggedright,
singlelinecheck=false
}
\caption{Comparison of diagnostic results between different background pressure, $P_{bg}$, for the $f_b=1.0$, $f=0.1$ equilibrium load: With $\tau=t/T_{ce}$ the set of plots are zoomed in showing their respective diagnostic readings upto $\tau=8500$ for clarity. $T_{bg}$ is the temperature of the background gas for all $P_{bg}>0$. (a) The radius of the centre-of-mass of the electron component, $R_{cme}$ normalised by the wall radius $R_W$ is plotted as a function of time. \textit{Inset}: The same set of plots is extended upto the end of the simulations at $\tau=22700$ (b) The left potential probe's reading, $V_p$ as a function of time at different background pressures. (c) The Potential Energy of the electron component, $J_e$ normalised by the initial total energy of the 2-component plasma, $E_0$ is plotted as a function of time.(d) The Potential Energy of the ion component, $J_i$ normalised by the initial total energy of the 2-component plasma, $E_0$ is plotted as a function of time. }
\end{figure*}

Davidson and Uhm's linear analytical theory\cite{rdavh1}, when applied on the first equilibrium ($f_b=0.02$, $f=0.15$) at zero background pressure (absence of neutrals) predicts growth of an unstable $l=1$ azimuthal mode ($l$ being the mode number) on both components with angular frequency $2.827\times10^6 rad/sec$ and growth rate $9.47\times10^5 rad/sec$. Our simulation of the same equilibrium sans neutrals are in excellent agreement with the linearised theory in the linear phase of the instability. We get an unstable $l=1$ mode with angular frequency $2.792\times10^6 rad/sec$ and growth rate $9.6\times10^5 rad/sec$. A linear perturbation analysis of the second equilibrium ($f_b=1.0$, $f=0.1$) at zero background pressure predicts that this equilibrium should be stable\cite{rdavh1}. However the corresponding simulation reveals that this equilibrium infact destabilizes very slowly in the the form of a nonlinear $l=1$ mode accompanied by a few higher weaker modes\cite{megh3}. A linearly fitted growth rate of the impure (mixed with a few higher weaker modes) $l=1$ mode  is obtained as $4.0\times10^5 rad/sec$ while its angular frequency comes out to be $2.733\times10^7 rad/sec$\cite{megh3}. The slower nonlinear destabilization of the second equilibrium in absence of neutrals can be attributed to nonlinear terms in the perturbed equations of motion that have been neglected in its linear analysis. That the destabilization of the second equilibrium is not a numerical artefact has been verified in our previous work\cite{megh3} by repeating the experiment with greater number of pseudo particles for both components. The repeated simulations also gave results in perfect agreement with original simulation at zero background pressure.

We now subjected these two unstable equilibria to three different values of neutral background pressures, $P_{bg}$. The values of $P_{bg}$ chosen are $2\times10^{-8} Torr$, $2\times10^{-7} Torr$, and $2\times10^{-6} Torr$ at a common fixed background temperature $T_{bg}=300 K$. The collision time step, ${\Delta}t_c$ for the lowest background pressure was chosen as $6\times10^{-6} sec$ and it was reduced by an order of magnitude and two orders of magnitude for the $2\times10^{-7} Torr$ and $2\times10^{-6} Torr$ background pressures respectively. This kind of scaling of ${\Delta}t_c$ between the three background pressures helped in maintaining uniformity of the MCC routine's collision resolving capacity among the different experimental pressures. Ofcourse all the three values of ${\Delta}t_c$ satisfy the two criteria for collision time interval in their corresponding experiments. With the help of these PIC-with-MCC simulations we were able to study the effect of electron-neutral elastic (and exciting) collisions on two ion resonance instabilities that are growing via the fundamental $l=1$ mode at different rates. A point to be noted here is that in these sets of experiments the loaded ions are not of the same species as the background gas while in typical electron plasma traps the destabilizing ions are created from the ionization of the background atoms by electrons. Simulations of the latter kind is beyond the scope of this paper and has been reserved for our next work\cite{megh4}. For the present work we chose the ion species H+ for which the ion resonance instability is a theoretically and numerically well understood phenomenon\cite{rdav1,rdavh1,megh3}. The background gas with which the electrons make non-ionizing collisions is Argon. In our upcoming work we will be studying the effects of electron impact ionization of the Argon atoms on the dynamics of the plasma\cite{megh4}. As there are no ionizing collisions in the present set of numerical experiments the neutrals will only influence the instability through collisional relaxation of the electron cloud's profile. So the experiments of this Section can be described as a qualitative investigation of how such a process of profile relaxation of the electron cloud due to an arbitrary neutral species influences the ion-resonance instability due to an arbitrary ion contaminant mixed in the electron cloud.

The detailed linear and nonlinear dynamics of the first ($f_b=0.02$, $f=0.15$) equilibrium in the absence of any colliding neutrals has been described in detail in Sec. IV - subsection A of Reference 3 along with corresponding snapshots of the two-component plasma in course of the simulation (Fig 1 of Reference 3)\cite{megh3}. In brief, this unstable equilibrium leads to an in-phase excitation of the $l=1$ mode on both components of the plasma. The two components gradually get separated azimuthally (growing phase separation) as the orbital radius of the $l=1$ mode increases for both components. During the nonlinear growth phase of the mode, ions gain potential energy and are also collisionlessly heated by the instability. The gain in potential energy and kinetic energy of the ion component comes at the expense of the potential energy of electrons. The heated ion component develops a filamentation structure on its surface which eventually makes contact with the grounded wall during the unstable $l=1$ orbital motion. Soon after the entire ion population gets transported and lost to the grounded wall through this filamentation structure. The $l=1$ orbit of the electron component however stabilizes very close to grounded wall with the electron component profile deforming into an elliptical shape. Hence the final state of the ion resonance instability for this particular unstable equilibrium is an elliptical pure electron cloud making stable orbits very close to the grounded wall. 

Now we load this equilibrium in the presence of an Argon neutral background at the three experimental background pressures. For the three PIC-with-MCC simulations the electrons are also loaded with an added axial velocity component enabling them for 3D elastic and exciting collisions with background neutrals. What we observe from our diagnostics comprising of potential probes (Fig 1), radius of centre-of-mass of the electron component, and the potential energy of the electron component, is that there is negligible variation in the dynamics of this instability brought about by the electron-neutral non ionizing collisions for the above set of background pressures. The diagnostics record values with progression of the simulation that are identical to the zero pressure experiment for all the three experimental background pressures. More specifically for all the experimental  $P_{bg}$ values including $P_{bg}=0$, the potential probes record a similar exponential growth and saturation of $l=1$ mode, the potential energy of the electron cloud decreases and saturates similarly with time, and the radial location of the electron cloud also increases and saturates similarly with time. As a illustrative example, we have shown in Fig. 1 the left potential probe's reading for the lowest and the highest background pressures ($P_{bg}=0$ and $P_{bg}=2\times10^{-6} Torr$). It can bee seen that the the two signals lie nearly on top of one another. All these diagnostic results bring us to the conclusion that for this particular equilibrium at the chosen set of background pressures the electron-neutral elastic and exciting collisions can not influence the evolution of the ion resonance instability on the e- H+ cloud. We will get back to reason for this in a later comparison with the results for the second equilibrium.

Next we come to the set of experiments performed at $f_b=1.0$, $f=0.1$. Loading this equilibrium at zero pressure produces a slower growing $l=1$ mode than the $f_b=0.02$, $f=0.15$ equilibrium. Fig. 2 is a set of snapshots showing the evolution of the 2-component plasma profile for this unstable equilibrium at zero background pressure. The snapshots show that the $l=1$ mode is eventually excited after the plasma profile goes through a few higher weaker modes (the $l=3$ mode being the most visible in Fig. 2d). The $l=1$ mode gradually takes over as the most dominant mode (Fig. 2h-i) and towards the end of the simulation it is the only excited mode on the profile executing stable orbits (Fig. 2j-l) after having saturated its growth. We analysed the growth and frequency of the $l=1$ mode using our potential probes. Fig. 3a shows the signal recorded by the left potential probe for this simulation at zero background pressure. As explained earlier for this particular equilibrium the excited $l=1$ mode is nonlinear and mixed with other modes from the outset. However it shows a slow growth rate right up to its saturation phase (Fig. 3b). So it is worthwhile to get an estimate of the growth of this impure $l=1$ mode with a linear exponential fit on the growing part of the potential probe signal (Fig. 3c). This growth rate comes out to be $4.0\times10^5 rad/sec$ . We also performed a FFT on the truncated reading of Fig. 3b to obtain an estimate of the angular frequency of the mode which comes to be $2.733\times10^7 rad/sec$ (Fig. 3d). The energetics of the simulation at zero background pressure is shown in Fig 4. Here we can see that the nonlinear growth of the mode is associated with collisionless heating of the electrons by the instability at the expense of their potential energy. A smaller fraction of the potential energy of electron component is also pumped into the potential energy of the ion component during the growth of the mode. The energy exchange process stops with the saturation of the mode.

Now the $f_b=1.0$, $f=0.1$ equilibrium is loaded at the same experimental background pressures as the $f_b=0.02$, $f=0.15$ equilibrium. The electron now have cylindrical trap like axial velocities and are capable of 3D elastic and excitation collisions with the background neutrals. The collisions relax the profile of the electron cloud and the resultant changes in the electron density profile influences the ongoing collisionless ion resonance instability. Specifically collisional relaxation of the electron cloud reduces the average density of the electron cloud and thereby increases the fractional density of the ions mixed in it. The dynamically changing electron density and fractional density of ions feed back on the ongoing ion-resonance (two-stream) instability between the two components of the nonneutral cloud and produce deviations in the paths of progression of the instability that are uncorrelated at different background gas pressures. As the variations in the path of evolution of the plasma at the experimental background pressures  are small in comparison to average evolution over time it is sufficient to identify and discuss these variations only qualitatively without quantifying their magnitudes with direct measurements. We will discuss these variations with the help of the diagnostic results of Fig. 5. 

First let us look at radial location of the electron component as a function of time which directly shows how the $l=1$ mode grows in orbital radius and saturates at different background pressures for the electron component (Fig. 5a). It can be seen from Fig. 5a that the $P_{bg}=0$ load has the slowest increase in orbital radius of the electron cloud with the growth rate increasing the in the order of $P_{bg}=2\times10^{-7} Torr$, $P_{bg}=2\times10^{-6} Torr$, and $P_{bg}=2\times10^{-8} Torr$. After the growth period the $l=1$ mode for the lower background pressures of $P_{bg}=0$ and $P_{bg}=2\times10^{-8} Torr$ saturate at slightly smaller orbital radii as compared to the two higher background pressures of $P_{bg}=2\times10^{-7} Torr$ and $P_{bg}=2\times10^{-6} Torr$ (see inset of Fig. 5a). The potential probe readings also mirror the same information about the growth of the mode (Fig. 5b). 

Next we have also compared the time evolution of two energy components between the different background pressures, viz the potential energy energy of the electron cloud and the potential energy of the ion cloud. We have already seen in Fig. 4 that for the $P_{bg}=0$ load the potential energy of the electron component decreases while that of ion component increases during the growth of the instability. Fig. 5c and 5d show the potential energies of the electron component and ion component respectively upto the the growth phase of the instability for the experimental background pressures. It can be seen from Fig. 5c-d that the rates of decrease/increase in potential energy of electrons/ions increases with increase in background pressure. 

Hence we can conclude that changing the collisionality of the electron cloud by changing the background gas densities has brought about interesting subtle variations in the dynamics of the ion-resonance instability that are unique to each experimental background gas pressure. It is also to be noted that once the growth of the collisionless ion resonance instability saturates there is no further destabilization caused by the continuing collisions (see inset of Fig. 5a) which further emphasizes that the collisions are only influencing the dynamics of the cloud through feedback of electron cloud's profile relaxation on the growing ion resonance instability.

We have seen that elastic and excitation collision of electrons with background neutrals influence the dynamics of the ion resonance instability for the $f_b=1.0$, $f=0.1$ equilibrium while they have no effect on the instability dynamics for the $f_b=0.02$, $f=0.15$ equilibrium. From the nature of the growth of instability in the two equilibria it is evident that this is a due to the different lengths of growth phase in the two equilibria. The excited $l=1$ mode in the $f_b=0.02$, $f=0.15$ equilibrium grows faster and saturates quicker (Fig. 1) than the $f_b=1.0$, $f=0.1$ equilibrium (Fig. 3a). So at a particular background pressure there will be more elastic and excitation collisions occurring during the growth phase of the instability for the $f_b=1.0$, $f=0.1$ equilibrium as compared to the $f_b=0.02$, $f=0.15$ equilibrium. Hence there is more feedback on the instability by the collisional relaxation of the electron cloud for the $f_b=1.0$, $f=0.1$ equilibrium. The feedback process brings about the subtle changes in the evolution of this equilibrium at the experimental background pressures. Furthermore we have already seen that once saturation of the mode is achieved, the ongoing collisions  can not further destabilize the cloud. Hence a smaller time window of influence of the electron-neutral collisions for the $f_b=0.02$, $f=0.15$ equilibrium is the precise reason why we do not observe any deviation in dynamics at different experimental background pressures for this equilibrium.

\section{Dynamics of $l=1$ perturbed pure electron clouds in presence of collisional background}

\begin{figure}
\centering
\includegraphics[scale=0.7]{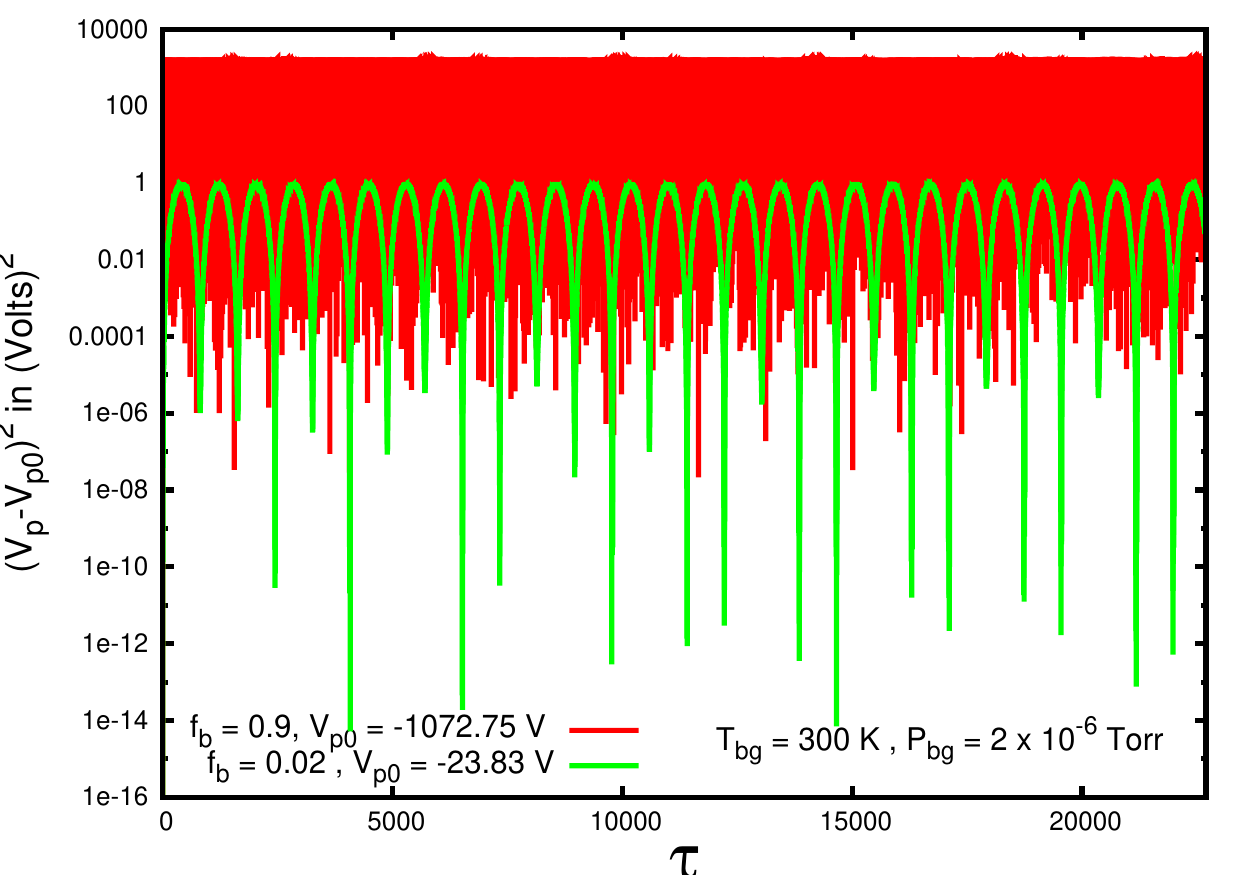}
\captionsetup{justification=raggedright,
singlelinecheck=false
}
  \caption{Potential probe data for the perturbed load (off-axis by $0.1\times{R_W}$) of the $f_b=0.02$ and the$f_b=0.9$ pure electron clouds at a common background pressure, $P_{bg}=2\times10^{-6} Torr$: $V_p$ is the reading of the left potential probe and $V_{p0}$ represents its corresponding initial value in the two simulations. The y axis is plotted in logscale for clarity. $T_{bg}=300 K$ is the temperature of the background gas in this set of runs. Normalized time, $\tau$ is in units of electron cyclotron time, \textit{i.e.} $\tau=t/T_{ce}$. Hence it is clear from the potential probe's data that there is no growth of the initially implanted $l=1$ mode in the simulation. The other three potential probes also gave similar results for this set of simulations.} 
\end{figure} 

In Sec III we have explained that the elastic and exciting collisions can influence the dynamics of the plasma through a feed back of the profile relaxation on the ongoing instability. But they can not by themselves destabilize the stable orbital motion of the plasma once saturation of the instability is reached. In this section we will put this reasoning to the test and verify that these collisions are indeed incapable of exciting an instability on the plasma profile. For this purpose we loaded two uniform density cylindrical pure electron profiles with a very small $l=1$ perturbation seeded from the outset \textit{i.e.} the profiles' central axis were shifted from the trap's central axis by a very small distance of $0.1\times{R_W}$. The Brillouin ratios of the profiles chosen were $f_b=0.9$ and $f_b=0.02$. It is well understood that in the absence of any other influence these profiles will continue to make stable $l=1$ obits at their preset orbital radius with their characteristic $l=1$ mode frequency\cite{rdav1,megh}. In our simulations we loaded these profiles at the highest experimental background pressure used in Sec III \textit{i.e.} $P_{bg}=2\times10^{-6} Torr$ to ensure maximum exposure of the profiles to elastic and excitation collisions during the simulations. However as expected there was no destabilization of the stable orbital motion caused by the collisions for both the high and the low Brillouin ratios. Fig. 6 shows the squared and logscaled plot of the reading of the left potential probe for these two simulations. The signals have been plotted in such a manner for accurate exponential fitting of any observable growth of the signals (see also Fig. 3d) and also for the purpose of depicting on the same graph the two signals with vast difference in their mean values as well as their amplitudes. Fig. 6 shows that both the signals remain perfectly stable throughout the simulations.

Our simulation results also contradict a theoretical model by Davidson and Chao\cite{chao} for the effects of electron-neutral elastic collisions on the dynamics of the cloud. This model predicts that the non-ionzing collisions should destabilize the initially stable $l=1$ orbital motion in the above two simulations into a growing $l=1$ mode with growth rate, $\gamma=2{\pi}{\nu_{en}}{\omega_{0}}/{\omega_{ce}}$\cite{chao}, where ${\nu_{en}}$, ${\omega_{ce}}$, and ${\omega_{0}}$  are the non-ionizing collision frequency in $s^{-1}$, cyclotron frequency in $rad/s$ and characteristic frequency of the $l=1$ mode\cite{rdav1} in the loaded electron cloud expressed in $rad/s$. The actual theory considered only electron-neutral elastic collisions contributing to the collision frequency ${\nu_{en}}$ . In applying the model to our numerical experiments we have suitably included the effectively elastic-like excitation collisions along with the regular elastic collisions in estimating an experimental value of ${\nu_{en}}$. The theoretical model\cite{chao} also assumed a constant temperature of the plasma profile maintained by electron-electron collisions in the derivation of the above expression for the growth rate, $\gamma$ due to electron-neutral collisions. We do not have electron-electron collisions happening in our simulations. But the plasma is loaded at zero temperature and does not undergo any collsionless heating process, although there is some amount of heating by the non ionizing electron-neutral collisions. So applying the above expression for $\gamma$ to our numerical experiments will include the approximation of a constant temperature of the plasma profile.      

The procedure for obtaining the growth rates from Davidson of Chao's\cite{chao} model when applied to our numerical experiments is as follows: From the initial cold loads of the plasma profiles it is simple to find an average velocity magnitude of electrons which we approximate as the average velocity magnitude all through the simulation. The total non-ionizing (elastic + excitation) collision cross-section can then be estimated using this average velocity magnitude, which leads to a value for the collision frequency ${\nu_{en}}$.  This value comes out to be $2.3985\times10^4$ per sec for the $f_b=0.02$ load and $1.7548\times10^4$ per sec for the $f_b=0.9$ load. Plugging in the other fixed frequencies in the above expression for the growth rate, $\gamma$, we can estimate the growth rate from Davidson and Chao's model to be $1.8838\times10^2$ rad/sec for the $f_b=0.02$ load and $6.2021\times10^3$ rad/sec  for the $f_b=0.9$ load. 

These growth rates imply that at the end of the simulations period of $5.4\times10^{-5}$ seconds, the $l=1$ orbital radius should have increased by $1.02\%$ for the $f_b=0.02$ load and by $39.78\%$ for the $f_b=0.9$ load according to the model of Davidson and Chao\cite{chao}. Even if we neglect the theoretical small increase in orbital radius for the $f_b=0.02$ load, the theoretically predicted increase in orbtial radius is substantial for the $f_b=0.9$ load and should be detected by the simulation diagnostics. However as we have discussed earlier using the potential probe diagnostic (Fig 6) the stability and initial amplitude of the $l=1$ mode remains completely unaltered in both the simulations, $f_b=0.02$ and $f_b=0.9$. A new theoretical explanation in support of our experimental results and their contradiction with the model given by Davidson and Chao\cite{chao} is provided below.   
 
The electron plasma in diocotron motion, \textit{e.g.} a stable $l=1$ orbit, can be thought of as having its electrostatic potential energy in two parts. The first part is the electrostatic energy of the profile without the surface wave/ diocotron mode and the other part which is a negative quantity is the electrostatic energy associated with the diocotron mode. Any process that further draws out energy from the mode \textit{i.e.} makes the mode's electrostatic energy even more negative and thereby also reduces the total potential energy of the profile,  will bring about growth of the mode. A process that only reduces the first part of the potential energy \textit{i.e.} the profile's electrostatic energy by virtue of its internal configuration without any surface wave, can not cause growth of the dioctron mode. Through our simulations we found that the elastic and excitation collisions can only reduce this first part of the potential energy by means of profile relaxation. However these collisions can not directly influence the energy associated with diocotron mode. On the other hand a process like a resistive wall instability can directly takes away energy from the diocotron mode by changing surfaces charge configuration on the grounded wall (or alternatively the virtual image charge configuration) that interacts with the diocotron mode of the plasma. So while a resistive wall may cause growth of negative energy dioctron modes on the electron plasma the elastic and excitation collisions can not directly influence diocotron motion by virtue of their potential energy reduction capability. Hence contrary to the theory by Davidson and Chao\cite{chao} which predicts that elastic collisions can reduce both parts of the potential energy of an electron cloud and destabilize them, our simulations show that elastic (and excitation) collisions can only cause reduction in the first part of the potential energy through relaxation of the profile.  

\section{Validation of the MCC numerical scheme}

\begin{table*}
\captionsetup{justification=raggedright,
singlelinecheck=false
}
\caption{\label{tab:table1} Cartesian velocity components and kinetic energies of arbitrarily chosen colliding electrons (pseudo electrons which are scaled down in mass inside the MCC routine) just before and just after the numerical collision operation: $v_{x0}$ , $v_{y0}$, and $v_{z0}$ are three components of the electron's velocity just before the collision, with $v_{z0}$ being the axial directions. $v_{x1}$ , $v_{y1}$, and $v_{z1}$ are the same velocity components just after the collision. ${\epsilon}_{k0}$ and ${\epsilon}_{k1}$ represent the kinetic energy of the electron just before and just after the collision. The time of collision, $t_{col}$ and the type of collision, $CT$ - ({\bf EL})astic or ({\bf EX})citation are also tabulated.}

\begin{ruledtabular}
\begin{tabular}{cccccccccc}
$v_{x0} (m/s)$ & $v_{y0} (m/s)$  & $v_{z0} (m/s)$ & ${\epsilon}_{k0} (eV)$ & $t_{col} ({\mu}s)$  & $CT$  & $v_{x1} (m/s)$ &  $v_{y1} (m/s)$  &$v_{z1} (m/s)$ & ${\epsilon}_{k1} (eV)$\\ 
\hline
\\

\tiny{$-20251700.4777294$} & \tiny{$1175073.05198364$}  & \tiny{$-10272700.0000000$} & \tiny{$1469.84998176886$} & \tiny{$0.6$}  & \tiny{$EL$}  & \tiny{$-19516351.0169898$} &  \tiny{$2245227.82545585 $}  & \tiny{$-11450392.4387800$} & \tiny{$1469.84986180629$}\\

\\
\tiny{$-40854485.1073843$} & \tiny{$13575274.2347969$}  & \tiny{$10272700.0000000$} & \tiny{$5568.80333388360$} & \tiny{$27.6$}  & \tiny{$EL$}  & \tiny{$-39096793.4565210$} &  \tiny{$4628854.69327025$}  & \tiny{$20221704.5292680$} & \tiny{$ 5568.79622455448$}\\
\\
\tiny{$1316893.20280337$} & \tiny{$-26955331.0768088$}  & \tiny{$10494034.6948208$} & \tiny{$2383.55397402810 $} & \tiny{$45.6$}  & \tiny{$EL$}  & \tiny{$710343.788248104 $} &  \tiny{$-27151109.5770355$}  & \tiny{$10038002.6573024 $} & \tiny{$ 2383.55395005079$}\\
\\
\\
\tiny{$5140270.17678625$} & \tiny{$-13367586.5916318$}  & \tiny{$10272700.0000000$} & \tiny{$883.100541131483 $} & \tiny{$0.6$}  & \tiny{$EX$}  & \tiny{$7203428.86747464 $} &  \tiny{$-13903403.3192488 $}  & \tiny{$7834977.10927241  $} & \tiny{$871.550540940748$}\\
\\
\tiny{$365257.220980826$} & \tiny{$-4167743.83521545$}  & \tiny{$-10272700.0000000$} & \tiny{$349.756692587647$} & \tiny{$27.6$}  & \tiny{$EX$}  & \tiny{$-2560205.78911724 $} &  \tiny{$ -5110140.42010097 $}  & \tiny{$-9289835.85222759   $} & \tiny{$338.206692396912$}\\
\\
\tiny{$18744576.1081227$} & \tiny{$-19513541.1212450$}  & \tiny{$752752.461213745$} & \tiny{$2082.94122670152$} & \tiny{$45.6$}  & \tiny{$EX$}  & \tiny{$17956984.2621484 $} &  \tiny{$-20049772.6621182$}  & \tiny{$2048400.16132038$} & \tiny{$2071.39097050021$}\\

\end{tabular}
\end{ruledtabular}

\end{table*}  

In order to verify the correctness and accuracy of our simulation results in Sec. III and IV we provide here a statistical and numerical validation of the MCC routine employed in these numerical experiments. The PIC part of the code is already very well benchmarked\cite{megh, megh2, megh3}. For the MCC routine validation procedure we have picked the PIC-with-MCC simulation of the $f_b=1.0$ ,$f=0.1$ equilibrium at $P_{bg}=2\times10^{-7} Torr$ described in Sec III. 

We will now verify that the total number of elastic and excitation collisions simulated by the MCC routine in the simulation time period are correct numbers coherent with a real experiment with same initial conditions. For this purpose we will proceed to make a rough theoretical estimate of the number of collisions that should happen in course of the simulation. We already know that electron cloud is loaded with an equilibrium rigid rotation and a given initial magnitude of axial velocity. Taking the mean radial co-ordinate of electrons as half the radial extent of the cloud we can multiply this radius with the electron cloud's angular velocity to obtain an estimate of the average linear velocity of electrons in the plane of the PIC simulation. The magnitude of the average velocity in the plane of PIC simulation and the fixed initial magnitude of the axial component of velocity can be used to determine the average magnitude of resultant velocity of the loaded electron cloud. The initial average kinetic energy of electrons can also be calculated from this average speed of electrons. 

Next we make a simplifying assumption that average kinetic energy of electrons remains near about its calculated initial value all through the simulation, or atleast its order of magnitude remains unchanged (see the plot of the total kinetic energy of electrons as a function of time in Fig. 4). So the initial average kinetic energy of electrons can be taken as rough estimate of the average kinetic energy throughout the simulation. With this approximate value of average kinetic energy of electrons we calculated the average cross sections for elastic and excitation collisions of electrons with neutrals from the available analytical formulas for collision cross sections. Using these mean collision cross sections and the average speed of electrons calculated earlier the average probability of elastic and excitation collision of an electron in the collision interval is calculated for the given background pressure. These probabilities multiplied by the total number of real electrons considered in the simulation give the average number of elastic and excitation collisions occurring in the collision interval. A further multiplication by the total number of collision intervals in the simulation time period gives an estimate of the total number of elastic collisions and excitation collisions that occurs in the simulation time period. These numbers come out to be $1.158\times10^{12}$ elastic collisions and $1.130\times10^{11}$ excitation collisions. 

In our simulation we recorded the total number of elastic and excitation collisions for pseudo electrons in the simulation time period. Multiplying these numbers by the representation number for pseudo electrons gives us the total number of collisions by real electrons that were represented by the pseudo electron collisions. These numbers are obtained as $1.317\times10^{12}$ elastic collisions , $1.293\times10^{11}$ excitation collisions. So the collisions statics from simulation being in good agreement with corresponding theoretical estimates show that the MCC operates correctly in this aspect.

We will now test the correctness of the dynamics of collisions implemented by the MCC routine. For this purpose we randomly chose three elastic collisions and three excitation collisions occurring at different moments in course of the simulation. Recall that inside the MCC routine the pseudo electron is for all intents and purposes a real electron. Noting down the velocity components, and kinetic energy of the colliding electron prior to and after the collision in Table 1 we can check from the kinetic energy columns of Table 1 that for elastic collisions only a tiny fraction of the kinetic energy of the projectile electron gets transferred to the massive thermally moving neutral when they collide. For the excitation collision Table 1 shows that $11.55 eV$ of the projectile's kinetic energy is used up for excitation of the neutral and a tiny fraction of the remaining kinetic energy is transferred as kinetic energy to the neutral. Hence the dynamics of collisions are implemented correctly by the MCC. From the values of the velocity components before and after the collision we also see that electron is scattered by the collision which is the underlying cause of the profile relaxation induced by elastic and excitation collisions.  The common value of projectile axial velocity $v_{z0}=\pm1.02727\times10^7$ m/s for some of the collisions is because these particles did not undergo any collisions before this and hence still carry the common loaded magnitude of axial velocity (recall that axial velocities can only be altered by collisions and not by the PIC).

\section{Conclusions and discussions}

In this paper we have addressed how the dynamics of a trapped electron cloud is influenced by elastic collisions with an inert neutral gas that is always present in an experimental trap, usually at very low pressures. Our simulations reveal the true nature of the interaction of the  electron-neutral elastic collisions with the dynamics of the cloud. We have investigated how a ongoing collisionless ion-resonance instability due to ion impurities mixed with the cloud is dynamically influenced by the feedback of the electron cloud's collisional profile relaxation on the instability. The effect of feedback is visible in the energetics and growth rate of the instability. As this feedback is a highly dynamical nonlinear process it effects also do not follow any particular trends with increase/decrease of the background gas pressure. We also observed that the feedback can influence a growing ion resonance instability but can not alter the stable dynamics of the cloud once the instability has saturated. So if the period of growth of the instability is very short, then there may not be sufficient number of collisions in the short growth phase to influence the dynamics of the instability. Such a case of a quickly saturating ion resonance instability being unaltered by the elastic collisions was also shown in our simulations.  

We have also demonstrated that contrary to an existing theory\cite{chao} on influence of elastic collisions between electrons and background neutrals, the collisions themselves can not destabilize an otherwise stable electron cloud. This, we have shown, is because of the nature of extraction of potential energy from the cloud by the collisions.

Now that we have built some understanding of the effects of elastic and excitation collisions of trapped electrons with background neutrals, we will explore in our next work\cite{megh4}, how ionizing collisions of electrons with background neutrals together with elastic and excitation collisions influence the dynamics of a trapped electron cloud.

\begin{acknowledgements}

The computer simulations described in this paper were performed on the UDBHAV cluster maintained by the Computer Centre of IPR, and the DAE grid UTKARSH maintained by BARC, Mumbai. We would like to thank staff of both these offices for keeping these clusters up and running at all times.  

\end{acknowledgements}

\nocite{*}
\providecommand{\noopsort}[1]{}\providecommand{\singleletter}[1]{#1}%

\end{document}